\documentclass[conference]{IEEEtran}
\usepackage{amsmath,amssymb,amsfonts,mathrsfs,bm}
\usepackage{amstext}
\usepackage{upgreek}
\usepackage{multicol}
\usepackage{indentfirst}
\usepackage{graphicx}
\usepackage{paralist}
\usepackage{multirow}
\usepackage{tabularx}
\usepackage{enumitem}
\usepackage[noadjust]{cite}

\usepackage{booktabs}
\usepackage{subfigure}
\usepackage{color}
\usepackage{amsmath, bm}
\usepackage{makecell}

\IEEEoverridecommandlockouts
\allowdisplaybreaks
\usepackage{algorithm}
\usepackage{algorithmic}
\floatname{algorithm}{\small \bf Algorithm}

\usepackage{accents}

\newcommand{\tabincell}[2]{\begin{tabular}{@{}#1@{}}#2\end{tabular}}
\newcommand{\RNum}[1]{\uppercase\expandafter{\romannumeral #1\relax}}
\allowdisplaybreaks
\begin{document}
\title{Deep Reinforcement Learning with Symmetric Prior for Predictive Power Allocation to Mobile Users}
\author{
 	\IEEEauthorblockN{Jianyu Zhao, Chenyang Yang}\\
	\IEEEauthorblockA{Beihang University, Beijing, China\\
	Email: jianyuzhao$\_$buaa@163.com, cyyang@buaa.edu.cn}
}
\maketitle

\begin{abstract}
Deep reinforcement learning has been applied for a variety of wireless tasks, which is however known with high training and inference complexity. In this paper, we resort to deep deterministic policy gradient (DDPG) algorithm to optimize predictive power allocation among $K$ mobile users requesting video streaming, which minimizes the energy consumption of the network under the no-stalling constraint of each user. To reduce the sampling complexity and model size of the DDPG, we exploit a kind of symmetric prior  inherent in the actor and critic networks: permutation invariant and equivariant properties, to design the neural networks. Our analysis shows that the free model parameters of the DDPG can be compressed by $2/K^2$. Simulation results demonstrate that the episodes required by the learning model with the symmetric prior  to achieve the same performance as the vanilla policy reduces by about one third when $K=10$.
\end{abstract}

\section{Introduction}
Deep reinforcement learning (DRL) has a wide range of applications in wireless tasks \cite{DRL2019,zhang2019proactive,Jianj2019GC,gadaleta2017d}, aimed to make decision for resource management in an on-line, end-to-end, model-free or distributed manner.

One of the important applications is predictive resource allocation, which has been shown to provide a remarkable gain in terms of boosting the performance of mobile networks by optimizing radio resources based on future information \cite{mobility,LTEU,scy}. Most existing works optimize the resource allocation policy either assuming perfect future information \cite{scy} or using the predicted information with machine learning \cite{mobility,LTEU}. By resorting to reinforcement learning, the policy can be optimized directly from current and past observations, as illustrated in \cite{LDGC19} by designing an energy-saving policy for video transmission under the quality-of-service (QoS) constraint of a mobile user.

Nonetheless, DRL-based solutions are known with high sample complexity, i.e., the number of episodes required to achieve a desirable performance is prohibitively large. For example, the policy derived in \cite{LDGC19} converges to the optimal policy after $5 \times 10^4$ episodes, which is about 2000 hours with each episode nearly 150 s! This is unacceptable for the practical use of most wireless applications.

One possible approach to improve sample efficiency is to introduce inductive bias into the structure of deep neural networks (DNNs) \cite{fastandslow}. By exploiting the priori knowledge for the input-output relation underlying a task, the hypothesis space for searching the model parameters of a DNN can be reduced, such that fewer samples are required for  training.
As a large class of priori knowledge, permutation invariance (PI) or permutation equivalence (PE) has been embedded into DNN by parameter sharing to reduce the sample complexity \cite{SEquivariance}. A natural question is: can we harness this class of priors for DRL to reduce the required episodes?

In this paper, we make an attempt to reduce the sample complexity of DRL. In particular, we resort to deep deterministic policy gradient (DDPG) algorithm \cite{DDPG} to optimize  predictive power allocation for video streaming that minimizes the average energy consumed by base stations (BSs) to ensure the QoS of every mobile user. By revealing and leveraging the PI and PE properties inherent in the actor and critic networks of DDPG, we design a PE/PI-DDPG by introducing parameter sharing into fully-connected DNNs (FC-DNNS). Our results show that the numbers of episodes required for convergence and the free model parameters for training the PE/PI-DDPG are much less than the FC-DNN based DDPG when the number of users is large.

\section{System Model and Problem Formulation}
Consider a learning-enabled cellular network, where $M$ BSs connected with a central unit (CU) serve $K$ mobile users. The CU monitors and records the status of each user via the BSs, learns a resource allocation policy for the users, and controls the BSs to execute the policy by sending instructions. The users move across multiple cells during video streaming. We assume that  each user is associated with the BS that has the strongest large-scale channel gain.

Each video is divided into $N_v$ segments. The playback duration of each segment is divided into $N_f$ time frames, each with duration $\Delta T$. Assume that the large-scale channel gains are constant in each frame but may change among frames. Each frame is further divided into $N_s$ time slots, each with duration $\tau$, i.e., $\tau = \Delta T/N_s $. Assume that the small-scale channel gains remain constant in each time slot and are independently and identically distributed among time slots.

Denote the large-scale channel gain from the $k$th user to its associated BS in the $t$th frame as $\alpha_t^k$, and the small-scale channel gain from the user to the BS in the $j$th time slot of the $t$th frame as $g_{tj}^k$. When the users are served with orthogonal frequency division access, the data rate of the $k$th user in the $j$th time slot of the $t$th frame can be expressed as
$R_{tj}^k = W^k \log_2 \left(1 + \frac{\alpha_t^k g_{tj}^k}{\sigma_0^2} p_{tj}^k \right)$,
where $W^k$ is the bandwidth for the $k$th user, $p_{tj}^k$ is the transmit power allocated to the $k$th user in the $j$th time slot of the $t$th frame, and $\sigma_0^2$ is the noise power.

To avoid stalling, each video segment should be delivered to the buffer of each user before playback. Denote $S_n^k$ (in bits) as the size of the $n$th segment in the video requested by the $k$th user, then the QoS constraint of the user can be expressed as
$\sum_{n=1}^{l} \sum_{t=(n-1)N_f + 1}^{nN_f}  \sum_{j=1}^{N_s} \tau R_{tj}^k\geq \sum_{n=2}^{l+1}  S_n^k$, $l = 1,\cdots,N_v -1$.
The total energy consumed at the BSs by delivering the video to the $k$th user in the $t$th time frame is
$E_t^k = \frac{1}{\rho} \sum_{j=1}^{N_s} \tau p_{tj}^k + \Delta T P_c$,
where $\rho$ reflects the impact of power amplifier, cooling and power supply, and $P_c$ is the power for operating the baseband and radio frequency circuits.

We optimize power allocation among the users to minimize the average total energy consumed at the BSs required to ensure the QoS of every mobile user, i.e.,
\begin{subequations} \label{eqn:p0}
	\begin{align}
	\min_{\{p_{tj}^k\}}~& \mathbb{E} \left[\sum_{k=1}^{K} \left(\sum_{t=1}^{N_vN_f} \left(\frac{1}{\rho} \sum_{j=1}^{N_s} \tau p_{tj}^k + \Delta T P_c \right)\right)\right] \label{eqn:obj}\\
	s.t.~ &\sum_{n=1}^{l} \sum_{t=(n-1)N_f + 1}^{nN_f}  \sum_{j=1}^{N_s} \tau R_{tj}^k\geq \sum_{n=2}^{l+1}  S_n^k, \label{eqn:c1}\\
	& ~ l = 1,\cdots,N_v -1, k = 1,\cdots,K \nonumber  \\
	& \sum_{k=1}^{K} p_{tj}^k I(m,k) \leq P_{\max} , \quad m=1, \cdots, M \label{eqn:c2}
	\end{align}
\end{subequations}
where the average is taken over both large-scale and small-scale channel gains, $I(m,k)$ is an indicator function with $I(m,k)$ = 1 if the $k$th user is associated with the $m$th BS, and $I(m,k)$ = 0 otherwise.

At the time instance when a user initiates video streaming, i.e., the start of the first time slot in the first frame, the future values of $\alpha_t^k$ and $g_{tj}^k, t=1,\cdots, N_vN_f, j=1,\cdots, N_vN_fNs$ are unknown. To find the solution of the problem in \eqref{eqn:p0}, we resort to DRL to make the prediction and optimization simultaneously from observations in an end-to-end manner.

\section{DRL-based Predictive Power Allocation}

In a standard RL framework, the agent observes a state $s_t$ from environment at time step $t$ and selects an action $a_t$ based on a policy $\pi$. One time step later, the agent receives a reward $r_{t+1}$ as the consequence of the action, and observes a new state $s_{t+1}$. The goal of the agent is to find a policy $\pi^*$ that maximizes the expected return starting from $s$, i.e., the state-value function $V_{\pi}\left(s\right)=\mathbb{E}\left[\sum_{t=0}^{T}\gamma^{t} r \left(s_t,\pi\left(s_t\right) \right)|s_0=s \right]$, where $\gamma$ denotes the discount factor.

\subsection{Formulating Problem \eqref{eqn:p0} in RL Framework}
The power allocation optimization can be formulated as the following RL problem, where the CU serves as the agent.

\subsubsection{Action}
 A straightforward formulation is to regard $p_{tj}^k$ as the action, which however is harder to learn and incurs large signaling overhead between the CU and BSs \cite{LDGC19}. As analyzed in \cite{LDGC19}, optimizing $p_{tj}^k$ is equivalent to optimize the average rate $\bar R_t^k \triangleq \mathbb{E}_g[R_{tj}^k]$. In particular, a BS can adjust the transmit power in the $j$th time slot of the $t$th frame with the average rate according to $p_{tj}^k = p^{opt} \left( \alpha_t^{k}, g_{ti}^{k}, \bar R_t^{k} \right)$, which can be obtained from a water-filling power allocation policy and the relation between $\bar R_t^k$ and the water-level. Therefore, the {\bf action} vector is the average rate of all users in each frame,
\begin{equation}
\bm a_t = \left[ \bar R_t^{1}, \bar R_t^{2}, \cdots, \bar R_t^{K}\right]^\mathsf{T}
\end{equation}
where $[\cdot]^\mathsf{T}$ denotes transpose. Then, the duration of a time step in our RL formulation is equal to the frame duration.

\subsubsection{State}
To ensure the QoS and maximal power constraints, the buffer status and the associated BS of each user should be considered. Denote $B_t^{k}$ and $M_t^{k}$ as the amount of data remaining in the buffer of the $k$th user and the index of the BS the $k$th user associated with in time step $t$, respectively. Denote $l_{t}^k$ as the index of the frame of the video segment the $k$th user playback at time step $t$. Since $l_{t}^k$ reflects the playback progress of the current segment of the user and affects the transition of $B_t^{k}$ to $B_{t+1}^{k}$, it should be included into the state. Since the fraction of a video having been downloaded affects the termination of an episode, the ratio of the accumulatively downloaded bits to the whole video of the $k$th user,
$\eta_t^{k} = \frac{\sum_{i=1}^{t} \sum_{j=1}^{N_s} \tau R_{ij}^k }{\sum_{n=1}^{N_s} S_n^k}$,
is useful for the agent to make the optimization. Since the average energy consumption for video streaming depends on the large-scale channel gain, $\alpha_t^k$ should be an element of the state. To help the agent make the prediction, the large-scale channel gains in the past $N_t$ time steps are also included in the state. Since a user may have the same large-scale channel gain at different cells, $\bm \alpha_t^k \triangleq [\alpha_{1, t}^k, \cdots, \alpha_{N_b, t}^k]$  should be in the state, where $\alpha_{i,t}^k$ is the large-scale channel gain between the $k$th user and the $i$th neighbouring BS and $\alpha_{1,t}^k = \alpha_{t}^k$. Then, the state vector of the $k$th user can be expressed as $\bm s_t^k \triangleq [B_t^{k}, l_t^k, \eta_t^{k}, M_t^{k}, \bm \alpha_{t-N_t}^k, \cdots, \bm \alpha_t^k]$. Finally, the {\bf state} matrix is
\begin{equation}
\footnotesize
{\begin{matrix}
\bm s_t\!=\!\begin{pmatrix}
\bm s_t^1 \\
\vdots \\
\bm s_t^K
\end{pmatrix}\!=\!\begin{pmatrix}
B_t^{1} & l_t^1 & \eta_t^{1} & M_t^{1} & \bm \alpha_{t-N_t}^1 & \!\cdots\! & \bm \alpha_t^1 \\
\vdots & \vdots & \vdots & \vdots & \vdots & \quad & \vdots \\
B_t^{K} & l_t^K & \eta_t^{K} & M_t^{K} & \bm \alpha_{t-N_t}^K & \!\cdots\! & \bm \alpha_t^K\\
\end{pmatrix}
\end{matrix}}
\end{equation}
which has $K$ rows and $4+\left(N_t+1\right)N_b$ columns.

\subsubsection{Reward}
To ensure the constraints in \eqref{eqn:c1} and \eqref{eqn:c2}, one can employ the safe layer method  \cite{dalal2018safe}, where at time step $t = lN_f$, $l=1, \cdots,N_v-1$, action $\bm a_{t}$ is transformed into $\tilde{\bm a}_{t}$ by solving the following optimization problem,
\begin{subequations} \label{p1}
\begin{align}
	\min_{\tilde{\bm a}_{t}} ~& \left\|\bm a_{t}-\tilde{\bm a}_{t}\right\|_2  \label{eqn:obj1}
	\\ s.t.~ &\bm a_1 +\cdots + \bm a_{t-1} + \bm a_{t}   \geq \sum_{n=2}^{l+1} \bm S_n \label{eqn:c3}\\
	& \sum_{k=1}^{K} p^{opt} \left( \alpha_{lN_f}^{k}, g_{lN_fi}^{k},  \bar R_{t}^{k} \right)  I(m,k) \leq P_{\max} \label{eqn:c4} \\
	& m=1, \cdots, M \nonumber
	\end{align}
\end{subequations}
where $\bm S_n \triangleq \left(S_n^1, \cdots, S_n^K \right)$.
However, this optimization problem may not have feasible solution for the problem at hand. For example, if users are not in good channel conditions such that the amount of data in the buffer is much less than the amount of data required for playback at time step $t$, then the QoS cannot be ensured even when $P_{\max}$ is used.
To circumvent this difficulty, we introduce the safe layer only to ensure one constraint.

When solving the problem in \eqref{p1} only with the constraint in \eqref{eqn:c3} at time step $t$, both $\sum_{n=2}^{l+1} \bm S_n$ and $\sum_{i=1}^{t-1}  \bm a_t$ are fixed constants, thereby the solution can directly be obtained as $\tilde{\bm a}_{t} = \sum_{n=2}^{l+1} \bm S_n - \sum_{i=1}^{t-1}  \bm a_t$. However, the problem only with the constraint in \eqref{eqn:c4} does not have a simple solution due to the coupling of the users. Hence, we introduce a safe layer to satisfy the QoS constraint and impose a penalty on the reward when the maximal power constraint is not satisfied.

The {\bf reward} for the agent is designed as
\begin{equation}
\small
r_t = - \sum_{k=1}^{K} \sum_{j=1}^{N_s} \tau p_{tj}^k - \lambda \sum_{m=1}^M \sum_{k=1}^{K} \left(\sum_{j=1}^{N_s} \tau p_{tj}^k I(m,k)- P_{\max} \right)^{+}
\end{equation}
where $\left(x \right)^{+} = \max \left\{ x,0 \right\}$, $\lambda$ is the penalty coefficient, and $- \sum_{k=1}^{K} \sum_{j=1}^{N_s} \tau p_{tj}^k$ is the total transmit energy consumed by all users in the $t$th time step.
\subsection{Transmission Policy Based on DDPG}
Since the state matrix lies in continuous space,
 we resort to DDPG \cite{DDPG} to solve the problem \eqref{eqn:p0}.
DDPG maintains two DNNs  stored at the CU, namely actor network $\mu (\mathbf s;\bm \theta_{\mu})$ and critic network $Q(\mathbf s, \mathbf a; \bm \theta_Q)$. The actor network learns the \emph{policy function} (i.e., the mapping from the state matrix to the action vector), whose output is then used to compute $p_{tj}^k$. The critic network learns the \emph{action-value function} (i.e., $Q^{\pi}(\mathbf s, \mathbf a) \triangleq \mathbb E \left[\sum_{t=0}^{T}\gamma^{t} r \left(s_t,a_t \right)|s_t=\mathbf s, a_t=\mathbf a, \pi \right]$).

During the interactions with the environment, the CU collects the experience $\mathbf e_t =  [\mathbf s_t,  \mathbf a_t, r_t, \mathbf s_{t+1}]$ from the BSs in a database as $\mathcal{D}_\mathcal{B} = \{\mathbf e_1, \cdots, \mathbf e_t \}$.
At each iteration, a mini-batch of experience is sampled from the database $ \mathcal{D}_\mathcal{B}$ to update the model parameters, i.e., \emph{experience replay} \cite{mnih2015human}.

The model parameters of the critic network are updated with gradient descent as
\begin{equation}
\bm \theta_{Q} \leftarrow \bm \theta_{Q} - \frac{\delta_Q}{|\mathcal{B}|} \nabla_{\bm \theta_{Q}} \sum_{j\in \mathcal{B}} \left[ y_j - Q(\mathbf s_j, a_j;\bm  \theta_Q) \right]^2 \label{eqn:Q}
\end{equation}
where $y_j = r_j + \gamma Q'(\mathbf s_{j+1}, \mu'(\mathbf s_{j+1}; \bm \theta_{\mu}'); \bm \theta_{Q}')$, $\delta_Q$ is the learning rate of critic network, $Q'(\mathbf s, a; \bm\theta_{Q}')$ and $\mu'(\mathbf s;\bm  \theta_{\mu}')$ are the target critic network and target actor network, respectively, which have the same structure as $Q(\mathbf s, \mathbf a; \bm \theta_Q)$ and $\mu (\mathbf s;\bm \theta_{\mu})$, and are updated by $\bm \theta_Q' \leftarrow \bm \omega \theta_Q +  (1 - \omega) \bm \theta_{Q}'$ and $\bm \theta_\mu' \leftarrow \bm \omega \theta_\mu +  (1 - \omega) \bm \theta_\mu'$ with very small value of $\omega$ to stabilize the learning procedure \cite{DDPG}.

The model parameters of the actor network are updated using the sampled policy gradient as
\begin{equation}
\bm \theta_{\mu} \leftarrow \bm \theta_{\mu} +  \frac{\delta_\mu}{|\mathcal{B}|} \nabla_{a} Q(\mathbf s, a;\bm \theta_Q)|_{\mathbf s=\mathbf s_i, a = \mu(s_i;\bm \theta_\mu)} \nabla_{\bm \theta_\mu} \mu (\mathbf s|\bm \theta_{\mu})|_{\mathbf s = \mathbf s_i} \label{eqn:mu}
\end{equation}
where $\delta_\mu$ is the learning rate of the actor network.

We add a noise term $n_{t}$ sampled from a noise process $\mathcal N$ to the output of the actor network, i.e., $a_t = \mu(\mathbf s_t; \bm \theta_{\mu}) + n_t$, to facilitate the exploration.

\section{PE/PI-DDPG based Predictive Power Allocation}
In this section, we show that the policy function and the {action-value function} exhibit a kind of symmetric properties. We then design a PE/PI-DDPG, where the actor and critic networks are constructed by exploiting the properties.

\subsection{Permutation Equivariant and Invariant Properties}
We first define the PE and PI properties to be used in the sequel. Consider arbitrary permutation matrix $\bm \Pi$. A multivariate function $\bm y=f(\bm X)$ is one-dimension (1D)-permutation equivariant to $\bm X$ if $\bm \Pi \bm y=f(\bm \Pi \bm X)$, where $\bm X $ is a matric and $\bm y$ is a vector. A function $y=f(\bm X)$ is 1D-permutation invariant to $\bm X$ if $y = f(\bm \Pi \bm X)$, where $y$ is a scalar.

The policy function $\mathbf a = \tilde \mu (\mathbf s)$ maps the state of all users $\bm s = [\bm s^1, \bm s^2, \cdots, \bm s^K]^\mathsf{T}$ into the average rates of all users, i.e., $\bm a = \left[ \bar R^{1}, \bar R^{2}, \cdots, \bar R^{K}\right]^\mathsf{T}$. When the order of users in the state changes (say $\bm s^1$ swaps with $\bm s^2$), the order of average rates changes in the same way ($\bar R^{1}$ swaps with $\bar R^{2}$) but the policy function remains unchanged, i.e., $\bm \Pi\mathbf a = \tilde \mu (\bm \Pi\mathbf s)$. Hence, the policy function is 1D-permutation equivariant to the state.

The action-value function $q=\tilde Q(\mathbf s, \mathbf a)$  maps the state and action of all users into the expected return $q$.
By stacking the state and action according to the users' indexes,
\begin{equation}
{\begin{matrix}
\bm (\mathbf s, \mathbf a)=\begin{pmatrix}
B^{1} & l^1 & \eta^{1} & M^{1} & \cdots  & \bar R^{1}\\
\vdots & \vdots & \vdots & \vdots & \quad  &\vdots \\
B^{K} & l^K & \eta^{K} & M^{K} & \cdots &  \bar R^{K}\\
\end{pmatrix}
\end{matrix}}
\end{equation}
we can see that the expected return will remain unchanged if the order of the users changes. Hence, the action-value function is 1D-permutation invariant to $ (\mathbf s, \mathbf a)$.

\subsection{DDPG with PE/PI-DNNs}
By introducing parameter sharing into the FC-DNN, permutation equivariant DNN (PE-DNN) and permutation invariant DNN (PI-DNN) can be used to approximate the 1D-PE and the 1D-PI functions, respectively.

With the input as $\bm X \in \mathbb R^{n \times D}$, the weight matrix and the bias vector of the $i$th layer of PE-DNN can be respectively constructed as \cite{SEquivariance}
\begin{equation}
{\begin{matrix}
\bm W_i=\begin{pmatrix}
U_i & V_i & \cdots & V_i \\
V_i & U_i & \cdots & V_i \\
\vdots & \vdots & \ddots & \vdots \\
V_i & V_i & \cdots & U_i \\
\end{pmatrix}, &
\bm b_i = \begin{pmatrix}
P_i \\
P_i \\
\vdots \\
P_i \\
\end{pmatrix}
\end{matrix}}
\end{equation}
where $U_i$ and $V_i$ are sub-matrices, $P_i$ is a subvector. The number of columns of $U_i$ and $V_i$ is equal to $D$. The numbers of sub-matrices in each row and each column in weight matrix $\bm W_i$ are equal, which is identical to the number of subvectors in bias $\bm b_i$, and all are equal to $n$.

For PI-DNN with input $\bm X \in \mathbb R^{n \times D}$, the bias of the output layer is the same as FC-DNN and the weight matrix of the output layer can be constructed as
\begin{equation}
{\begin{matrix}
\bm W_i=\begin{pmatrix}
A_i & A_i & \cdots & A_i
\end{pmatrix}
\end{matrix}}
\end{equation}
where $A_i$ is a sub-matrix with the number of columns equal to $D$, and the number of the sub-matrices in the weight matrix $\bm W_i$ is equal to $n$. The input and all hidden layers are constructed the same as PE-DNN.

By designing the actor network as a PENN and the critic network as a PINN, the free model parameters that need to be trained can be computed as follows.
Since the number of free parameters in the bias is far less than that in the weight matrix, we only count the parameters in the weight matrix.

For notational simplicity, suppose that the number of layers is $H$ and the number of neurons in each hidden layer is $d$ for both actor and critic networks.

For the actor network, the dimension of the input is $K \times \left(4+\left(N_t+1\right)N_b\right)$, i.e., $n=K$ and $D=4+\left(N_t+1\right)N_b$, the dimension of the output is $K \times 1$. The free parameters in the weight matrix are in two sub-matrices, hence the number of parameters in the $i$th hidden layer can be expressed as $\frac{2d^2}{K^2}$, and the number of parameters in the input layer and the output layer can be expressed as $\frac{2dK\left(4+\left(N_t+1\right)N_b\right)}{K^2}$ and $\frac{2dK}{K^2}$, respectively. Consequently, the total number of model parameters can be obtained as $\frac{2\left(H-2\right) d^2 + 2dK\left(4+\left(N_t+1\right)N_b\right) + 2dK}{K^2}$.

For the critic network, the dimension of the input is $K \times \left(5+\left(N_t+1\right)N_b\right)$, i.e., $n=K$ and $D=5+\left(N_t+1\right)N_b$, the output is a scalar. The free parameters of the output layer in the weight matrix are in one sub-matrix, hence the number of parameters can be expressed as $\frac{d}{K}$, and the number of free parameters in the $i$th layer is the same as PE-DNN, for $1 \le i\le H-1$. Therefore, the total number of free parameters can be obtained as $\frac{2 \left(H-2\right) d^2 + 2dK\left(5+\left(N_t+1\right)N_b\right) + dK}{K^2}$.

If FC-DNNs are used, then the number of free parameters of the actor network is $ \left( H-2 \right) d^2 + dK\left(4+\left(N_t+1\right)N_b\right) + dK $, and the number of free parameters of the critic network is $\small \left( H-2 \right) d^2 + dK\left(5+\left(N_t+1\right)N_b\right) + d$. Hence, the actor and critic networks with parameter sharing can reduce the model parameters by $\frac{K^2\left(\left( H-2 \right) d^2 + dK\left(4+\left(N_t+1\right)N_b\right) + dK \right)}{2\left(H-2\right) d^2 + 2dK\left(4+\left(N_t+1\right)N_b\right) + 2dK}$ and $\frac{K^2\left(\left( H-2 \right) d^2 + dK\left(5+\left(N_t+1\right)N_b\right) + d \right)}{2 \left(H-2\right) d^2 + 2dK\left(5+\left(N_t+1\right)N_b\right) + dK}$ times with respect to the FC-DNN, respectively. When the width of the hidden layer $d$ is much larger than the dimensions of the input and the output  of the actor and critic networks, we can obtain
\begin{equation}
\footnotesize
\begin{split}
\frac{K^2\left(\left( H-2 \right) d^2 + dK\left(4+\left(N_t+1\right)N_b\right) + dK \right)}{2\left(H-2\right) d^2 + 2dK\left(4+\left(N_t+1\right)N_b\right) + 2dK} \\
\approx \frac{K^2\left(\left( H-2 \right) d^2 + dK\left(5+\left(N_t+1\right)N_b\right) + d \right)}{2 \left(H-2\right) d^2 + 2dK\left(5+\left(N_t+1\right)N_b\right) + dK} \approx \frac{K^2}{2} \label{eqn:para}
\end{split}
\end{equation}
This indicates that the number of model parameters in PE/PI-DDPG is $2/K^2$ of the DDPG based on FC-DNNs.

\section{Simulation Results}
In this section, we evaluate the performance of the PE/PI-DDPG by comparing with the DDPG based on FC-DNNs.

\subsection{Simulation Setup}
\begin{figure}[!htb]
	\centering
	\includegraphics[width=0.45\textwidth]{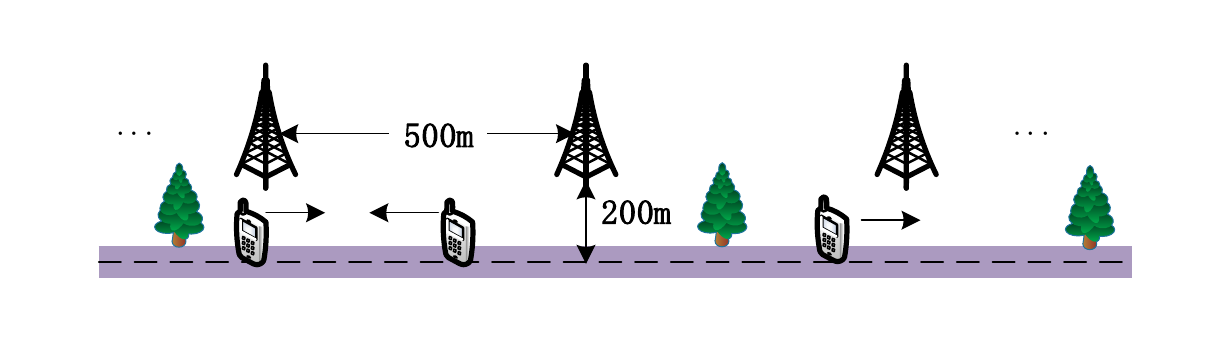}\vspace{-3mm}
	\caption{Simulation scenario.}
	\label{fig:layout}
\end{figure}

Consider a cellular network with multiple BSs located along a straight line, and $K$ users move along a straight road across cells, as shown in Fig. 1. The inter-BS distances are 500 m, and the minimum distance between the BSs and the road is 200 m. The maximal transmit power of each BS is 46 dBm. The noise power is -95 dBm/Hz and the bandwidth for each user is 2 MHz. Since the circuit energy consumption is identical for all the considered policies, we only consider transmit energy consumption. The path loss is modeled as $35.3+37.6\log_{10} (d)$ in dB, where $d$ is the distance between user and BS in meters. The small-scale channels follow Rayleigh fading. The playback duration of each video and each segment is 150 s and 10 s, respectively. Each segment is with size $1$ MBytes. Each time frame is with the duration of $\Delta T = 1$ s, and each time slot is with the duration of $\tau = 1$ ms, i.e., each frame contains $N_s = $ 1000 time slots. The user moves with random acceleration, where the acceleration in each frame is drawn from the Gaussian distribution with zero mean and standard deviation 0.5 m/s$^2$. The initial velocity of users is set as 16 m/s, and the minimal and maximal velocities of each user are 12 m/s and 20 m/s, respectively.

\subsection{Fine-Tuned Parameters for PE/PI-DDPG}
The actor network has four hidden layers each with 600 nodes, and employs a modified $\tanh$ function $0.5\times (\tanh(x) + 1)$ as the activation function in the output layer to bound the actions, where the upper bound is the average rate of the user in the best channel conditions on the road when $P_{\max}$ is used. The critic network first stacks the state and action together and then goes through four hidden layers each with 600 nodes, and  has no activation function in the output layer. All the hidden layers in the actor and critic networks use the rectified linear unit (ReLU) as the activation function. We use Adam \cite{adam} for learning the model parameters with a learning rate of 10$^{-4}$ and 10$^{-3}$ for the actor and critic networks, respectively. For the critic network, we include $L_{2}$ weight decay of 10$^{-4}$ to avoid over-fitting and use a discount factor of $\gamma = 1$.

We set $N_b = 2$ and $N_t = 2$ in the state. The update rate for the target networks is $\omega = 10^{-3}$. The replay memory size is $ |\mathcal{D}_\mathcal{B}| = 10 ^6$, and the mini-batch size for gradient descent is $|B|$ = 512. The penalty coefficient is set as $0.1$. The noise term $n_t$ follows Gaussian distribution with zero mean and the variance decreased linearly from $0.3$ to $0$.

\subsection{Performance Evaluation}
We compare the PE/PI-DDPG with the DDPG based on FC-DNNs from three aspects in terms of the total average energy consumption, the sample complexity, and the number of free parameters in DNNs.

\subsubsection{Average Energy Consumption} To evaluate the performance of the proposed policy, we compare the total average energy consumption of all users for each video achieved by the DDPG-based policies and the optimal policy, which is obtained by solving the problem \eqref{eqn:p0} assuming perfect prediction of large- and small-scale channel gains.
\begin{figure}[!htb]
	\centering
	\includegraphics[width=0.5\textwidth]{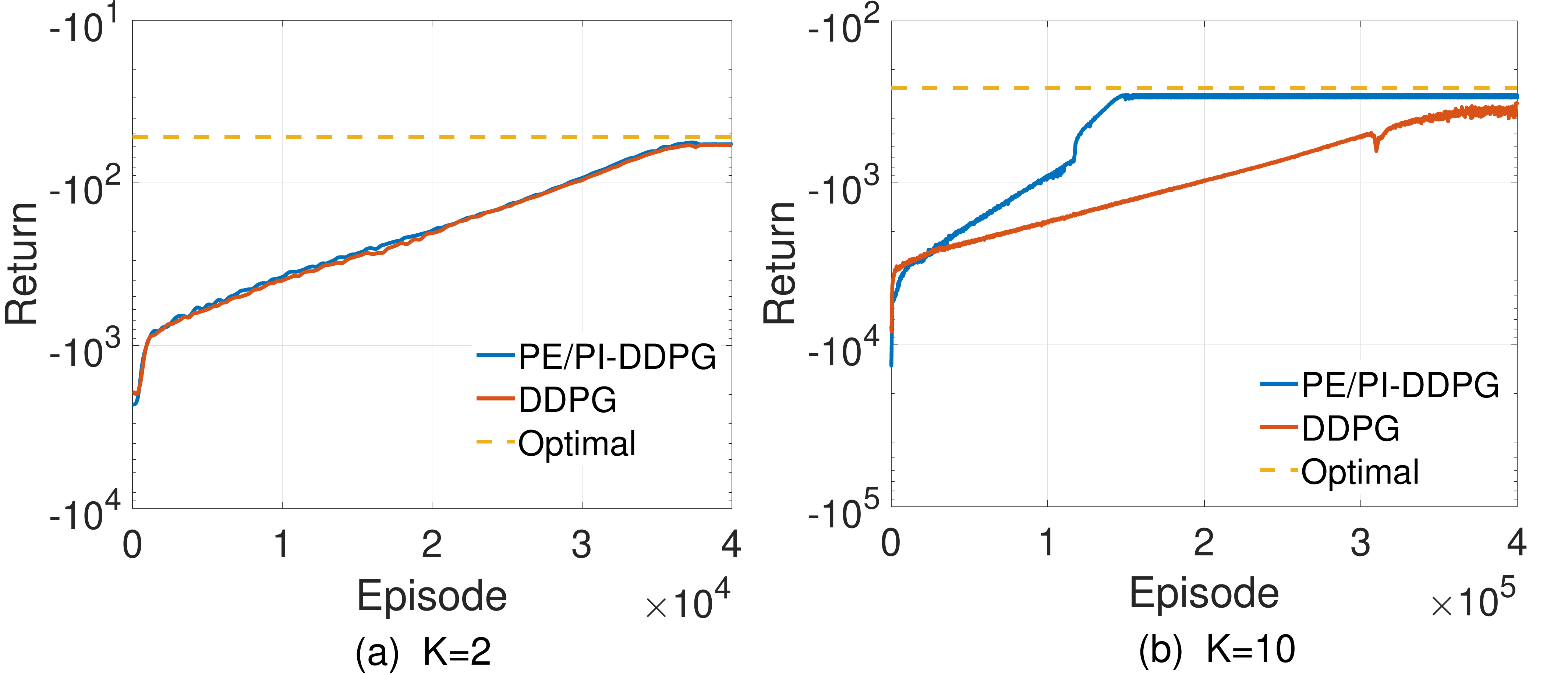}
	\vspace{-4mm}
	\caption{Return $\sim$ episodes, averaged over 10 Monte Carlo trials (each with random initial user locations and accelerations) over 400 successive episodes.}
	\label{fig:random}
\end{figure}
\vspace{-3mm}

In Fig. \ref{fig:random}, we show the learning curves of the PE/PI-DDPG and the DDPG-based on FC-DNNs (with legend ``DDPG"). Since there is no penalty in return after convergence, the negative of the converged return is the total energy consumed for all users. From Fig. \ref{fig:random}(a), we can see that both DDPG-based policies can converge to the optimal policy. From Fig. \ref{fig:random}(b), we can see that the PE/PI-DDPG approaches to the optimal policy much faster than the DDPG-based on FC-DNNs.

\subsubsection{Sample Complexity} The {sample complexity} of DRL is defined as the minimal number of episodes to achieve an expected performance  on the test set.

\begin{table}[!htbp]
        \caption{Sample Complexity  with different number of users.}\label{table}
    \centering
    \footnotesize
        \renewcommand\arraystretch{1.3}
    \begin{tabular}{c|c|c|c}
        \hline\hline
        {$K$}&PE/PI-DDPG& DDPG&{Compression ratio}\\
        \cline{1-4}
                \tabincell{c}
                {2}&{39,000}&{39,000}&{100$\%$} \\

        \cline{1-4}

        {5}&80,000&130,000&{61$\%$}\\

        \cline{1-4}

        {10}&135,000&400,000&{34$\%$}\\

        \hline\hline
    \end{tabular}
\end{table}
\vspace{-2mm}

In Table \RNum{1}, we compare the number of episodes required by PE/PI-DDPG to achieve the same expected return  with the DDPG with FC-DNNs (simply denoted as ``DDPG"). We can see that the compression ratio  increases with the number of users. When $K=10$, the sample complexity of the PE/PI-DDPG is much lower than the DDPG with FC-DNNs, i.e., the PE/PI-DDPG can converge three times faster.

\subsubsection{Number of Free Parameters} In Table \ref{table}, we provide the number of model parameters in the PE/PI-DDPG and the DDPG based on FC-DNNs  (again denoted as ``DDPG"). We can see that the compression ratio is $2/K^2$, which coincides with \eqref{eqn:para}. This indicates that the model size (and hence the memory to store the model for inference) of the PE/PI-DDPG is small for large number of users.

\begin{table}[!htbp]
        \caption{Number of free parameters with different number of users.}\label{table}
    \centering
    \footnotesize
        \renewcommand\arraystretch{1.4}
    \begin{tabular}{c|c|c|c}
        \hline\hline
        {$K$}&PE/PI-DDPG& DDPG&{Compression Ratio}\\
        \cline{1-4}
                \tabincell{c}
                {2}&{2,907,000 (2.9 M)}&{5,814,000 (5.8 M)}&{50$\%$} \\

        \cline{1-4}

        {5}&465,264 (0.46 M)&5,893,200 (5.9 M)&{8$\%$}\\

        \cline{1-4}
 
        {10}&116,376 (0.12 M)&6,025,200 (6.0 M)&{2$\%$}\\

        \hline\hline
    \end{tabular}
\end{table}
\vspace{0mm}

\section{Conclusion}
In this paper, we strived to reduce the sample complexity and model size of DRL-based policy by harnessing symmetric priors. We optimized predictive power allocation for video streaming over wireless networks to minimize the average energy consumption under the QoS constraint of every mobile user with DDPG. By exploiting the permutation invariant and equivariant properties in the actor and critic
networks, we constructed the two deep neural networks with parameter sharing. Simulation results showed that the number of episodes and the number of free model parameters required by the PI/PE-DDPG to achieve the same energy consumption of the DDPG based on FC-DNNs reduce quickly and even dramatically with the number of users. In fact, by designing the actor network as PI-DNN and the critic network as PE-DNN, the PI/PE-DDPG can adapt to the change of the number of users in addition to reducing the sample complexity and model size, which is important for mobile networks but will be evaluated in future.

\bibliographystyle{IEEEtran}
\bibliography{jianyubib}

\begin{thebibliography}{10}
\providecommand{\url}[1]{#1}
\csname url@samestyle\endcsname
\providecommand{\newblock}{\relax}
\providecommand{\bibinfo}[2]{#2}
\providecommand{\BIBentrySTDinterwordspacing}{\spaceskip=0pt\relax}
\providecommand{\BIBentryALTinterwordstretchfactor}{4}
\providecommand{\BIBentryALTinterwordspacing}{\spaceskip=\fontdimen2\font plus
\BIBentryALTinterwordstretchfactor\fontdimen3\font minus
  \fontdimen4\font\relax}
\providecommand{\BIBforeignlanguage}[2]{{%
\expandafter\ifx\csname l@#1\endcsname\relax
\typeout{** WARNING: IEEEtran.bst: No hyphenation pattern has been}%
\typeout{** loaded for the language `#1'. Using the pattern for}%
\typeout{** the default language instead.}%
\else
\language=\csname l@#1\endcsname
\fi
#2}}
\providecommand{\BIBdecl}{\relax}
\BIBdecl

\bibitem{DRL2019}
N.~C. {Luong}, D.~T. {Hoang}, S.~{Gong}, D.~{Niyato}, P.~{Wang}, Y.~{Liang},
  and D.~I. {Kim}, ``Applications of deep reinforcement learning in
  communications and networking: A survey,'' \emph{IEEE Commun. Surveys Tuts.},
  vol.~21, no.~4, pp. 3133--3174, Fourth quarter 2019.

\bibitem{zhang2019proactive}
Z.~{Zhang}, Y.~{Yang}, M.~{Hua}, C.~{Li}, Y.~{Huang}, and L.~{Yang},
  ``Proactive caching for vehicular multi-view {3D} video streaming via deep
  reinforcement learning,'' \emph{IEEE Trans. Wireless Commun.}, vol.~18,
  no.~5, pp. 2693--2706, May 2019.

\bibitem{Jianj2019GC}
J.~{Zhang}, Y.~{Huang}, J.~{Wang}, and X.~{You}, ``Intelligent beam training
  for millimeter-wave communications via deep reinforcement learning,'' in
  \emph{Proc. IEEE GLOBECOM}, 2019.

\bibitem{gadaleta2017d}
M.~Gadaleta, F.~Chiariotti, M.~Rossi, and A.~Zanella, ``D-{DASH}: A deep
  {Q}-learning framework for {DASH} video streaming,'' \emph{IEEE Trans. Cogn.
  Commun. and Netw.}, vol.~3, no.~4, pp. 703--718, 2017.

\bibitem{mobility}
N.~Bui and J.~Widmer, ``Data-driven evaluation of anticipatory networking in
  {LTE} networks,'' \emph{IEEE Trans. on Mobile Comput.}, vol.~17, no.~10, pp.
  2252--2265, Oct. 2018.

\bibitem{LTEU}
U.~{Challita}, L.~{Dong}, and W.~{Saad}, ``Proactive resource management for
  {LTE} in unlicensed spectrum: A deep learning perspective,'' \emph{IEEE
  Trans. Wireless Commun.}, vol.~17, no.~7, pp. 4674--4689, Jul. 2018.

\bibitem{scy}
C.{She} and C.{Yang}, ``Energy efficient resource allocation for hybrid
  services with future channel gains,'' \emph{IEEE Trans. Green Commun. and
  Netw.}, vol.~4, no.~1, pp. 165--179, Mar. 2020.

\bibitem{LDGC19}
D.~Liu, J.~Zhao, and C.~Yang, ``Energy-saving predictive video streaming with
  deep reinforcement learning,'' in \emph{Proc. IEEE GLOBECOM}, 2019.

\bibitem{fastandslow}
B.~Matthew, R.~Sam, X.~W. Jane, K.-N. Zeb, B.~Charles, and H.~Demis,
  ``Reinforcement learning, fast and slow,'' \emph{Trends Cogn. Sci.}, vol.~23,
  no.~5, pp. 408--422, 2019.

\bibitem{SEquivariance}
S.~{Ravanbakhsh}, J.~{Schneider}, and B.~{Poczos}, ``Equivariance through
  parameter-sharing,'' in \emph{Proc. JMCR ICML}, 2017.

\bibitem{DDPG}
T.~P. Lillicrap, J.~J. Hunt, A.~Pritzel, N.~Heess, T.~Erez, Y.~Tassa,
  D.~Silver, and D.~Wierstra, ``Continuous control with deep reinforcement
  learning,'' in \emph{Proc. ICLR}, 2015.

\bibitem{dalal2018safe}
\BIBentryALTinterwordspacing
G.~Dalal, K.~Dvijotham, M.~Vecerik, T.~Hester, C.~Paduraru, and Y.~Tassa,
  ``Safe exploration in continuous action spaces,'' \emph{arXiv preprint},
  2018. [Online]. Available: \url{http://arxiv.org/abs/1801.08757}
\BIBentrySTDinterwordspacing

\bibitem{mnih2015human}
V.~Mnih, K.~Kavukcuoglu, D.~Silver, A.~A. Rusu, J.~Veness, M.~G. Bellemare,
  A.~Graves, M.~Riedmiller, A.~K. Fidjeland, G.~Ostrovski \emph{et~al.},
  ``Human-level control through deep reinforcement learning,'' \emph{Nature},
  vol. 518, no. 7540, p. 529, Feb. 2015.

\bibitem{adam}
D.~P. Kingma and J.~Ba, ``Adam: A method for stochastic optimization,'' in
  \emph{Proc. ICLR}, 2014.

\end{thebibliography}

\end{document}